\documentclass[onecolumn,showpacs]{revtex4}

     \usepackage{graphicx}
     \usepackage{dcolumn}
     \usepackage{bm}
     \usepackage{epsf}

\begin{document}

\title{Two photon ionization of condensate atoms}

\author{M. Anderlini}
\altaffiliation[present address: ]{National Institute of Standards and Technology, Gaithersburg, MD 20899, USA.} %
 \author{E. Arimondo}
     \affiliation{INFM, Dipartimento di Fisica E. Fermi,
Universit\`{a} di Pisa, Largo Pontecorvo 3, I-56127 Pisa, Italy
     }%

     \date{\today}

     \begin{abstract}
The efficient photoionization of a Bose-Einstein condensate
requires the creation of ions with the smallest possible transfer
of atoms from the condensate into the thermal phase. The
spontaneous decay from excited states into atomic states with
momentum different from the initial one should be reduced. We
investigate theoretically the two-photon ionization of a rubidium
condensate using near resonant excitation to the 6P state and
second photon at 421 nm or 1002 nm into the continuum. Different
ionization schemes with coherent control of the first excitation
and reduction of the spontaneous decay are presented.
\end{abstract}

\maketitle

\section{Introduction}
The realization of Bose-Einstein condensates (BEC) of alkali atom
vapors has attracted much interest into new aspects of
photon-matter interaction arising from the quantum nature of the
atomic sample. Recently attention has been paid to the
photoionization of a BEC by monochromatic laser light
\cite{mazets,ciampini02,anderlini03,anderlini04,courtade04}. The
products of the photoionization process (electrons and ions) obey
to Fermi-Dirac statistics. Due to the quantum nature of the
initial atomic target and the narrow spectral width of the laser
ionization source, the occupation number of the electron/ion final
states may become close to unity, especially for a laser
excitation close to threshold. If this regime is reached the
ionization rate could be slowed down by the Pauli blockade, with
the occupation of ionized states determined by the balance between
the laser ionization of the condensate and the rate of escape of
the ionization products from the
condensed system. Moreover, the electrostatic  interaction between the ions and
neutral atoms in the condensate is expected to originate localized
deformations in the condensate density distribution. Either an
increase or a decrease in the neutral atom spatial density,
depending on the sign of the atom-ion phase shift, could
be originated in small regions surrounding the ions \cite{massignan05}. \\
\indent The experimental study of these processes requires the
production of a large ionic concentration within a condensate on
time shorter than the timescale on which the ions escape from the
system, of the order of hundred ns for typical condensate samples,
without simultaneously destroying the condensate. Instead, in the
two-photon ionization experiments reported in refs.
\cite{ciampini02,anderlini03,anderlini04,courtade04} the
photoionization laser produced a significant transfer of atoms
from the condensate into the thermal cloud. Those experimental
investigations demonstrated that the rubidium two-photon
ionization was not performed in efficient way, {\it i.e.} without
producing a large depletion of the condensate atoms with their
transformation into thermal atoms. Indeed, in typical two-photon
ionization experiments the ionization is increased only at the
expenses of a large production of thermal atoms. \\
\indent The present work reports a theoretical investigation of
the efficiency realized in the ionization of atomic rubidium. This
work addresses a target which is different from those associated
with the typical investigations of multiphoton processes. Our aim
is not to increase the multiphoton ionization probability per se.
In contrast, the production of ions is compared to optical pumping
rate, produced by the spontaneous emission, transferring
condensate atoms into the thermal cloud. The ionization {\it
efficiency} will be determined by the decrease in the loss towards
atomic states initially not occupied, i.e. a decrease of optical
pumping into sink thermal states, and by the increase in the
number of ions.\\
\indent Our condensate ionization scheme takes place in a three-level
cascade scheme, the final one being the atomic continuum. The
condensate losses are produced by spontaneous emission from the
intermediate excited state. An efficient ionization is realized
for an atomic transfer into the continuum with a reduced real
excitation of the intermediate level, whence a reduced role of
spontaneous emission processes from that intermediate level.
 The task of reducing the transfer towards
sink levels is usually reached applying the STIRAP technique for
the coherent transfer between the initial and final levels of a
three level scheme, as analyzed in ~\cite{bergmann,vitanov01}, and
for transfers to and from a continuum as in
\cite{vardi99,mackie00,vardi02}. An efficient STIRAP transfer,
which relies on the coherent coupling of the initial and the final
state, can be realized only using very short laser pulses, i. e.
shorter than the characteristic decoherence time of the system. In
the case of photoionization, the use of laser pulses shorter than
the typical electron-ion decoherence times would prevent from
creating the narrow kinetic energy distribution for the electrons
and ions required for the Pauli blockade. Instead we will analyze
here the two-photon ionization produced
by narrow band cw radiation. \\
\indent For long interaction times the photoionization and optical
pumping processes can be described through a simple rate equation
approach. At shorter interaction times, comparable to or smaller
than the spontaneous emission lifetime from the first excited
state, the determination of the photoionization efficiency
requires the solution of density matrix equations. Our solution of
the density matrix equations searching for a regime of efficient
ionization has strong connections with the investigations of refs.
\cite{vitanov02,halfmann} for the lineshapes of atomic excitation
and ionization in a cascade three-level system under two-color
excitation as a function of
the time delay in the application of the two lasers.\\
\indent Section II describes the characteristics of our three-level scheme.
 Section III
analyzes the case of simultaneous application of the excitation
and photoionization lasers for long interaction times on the basis
of the rate equation solution. In Section IV we analyze the
ionization efficiency for different temporal shapes of the
excitation and ionization laser pulses using the density matrix
approach. The lengths of the laser pulses and their time delay
represent the experimental parameters controlling the ionization
efficiency. Section V analyzes ionization based on a rapid
adiabatic transfer.

\begin{figure}[ht]
\centering
\includegraphics[scale=0.6]{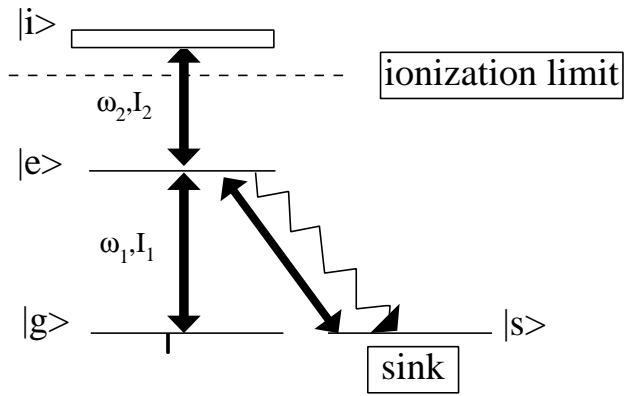}\\
\caption{Three-level scheme for the excitation from the condensate ground state to the
condensate ionic state through an intermediate $|e\rangle$ state. The
thermal sink state
$|s\rangle$ is occupied by the spontaneous emission decay from the
$|e\rangle$ state.}
\label{scheme}
\end{figure}

\section{atomic level and Hamiltonian}
We investigate the rubidium two-photon ionization scheme from the
$\mathrm{5S_{1/2}}$ ground state to the continuum with
intermediate resonant state the $\mathrm{6P_{3/2}}$ state at 2.94
eV above the $\mathrm{5S_{1/2}}$ ground state as in the experiment
of ref. \cite{anderlini04}. The two-photon ionization was produced
either by two photons a 421 nm or by a two color process, one
photon at 421 nm and a second one at 1002 nm. The photoionization
cross sections for absorption of one 421 nm or one 1002 nm photon
from the $\mathrm{6P}$ state have a small difference, not playing
an important role for the present investigation of the
photoionization efficiency. Instead a larger intensity was
available at the 1002 nm wavelength, and the contribution to
ionization of the two color process was much larger than the
contribution of the one-color photoionization by two 421 nm
photons.

The level scheme of Fig. 1 schematizes the two-photon ionization
process from the ground state $|g\rangle$ of the condensate to the
condensate ionic state $|i\rangle$ with a near resonant excitation of the
intermediate level $|e\rangle$. The population of the
intermediate state $|e\rangle$ may decay back into the ground state
through emission of spontaneous emission photons. For a sample
composed by a Bose-Einstein condensate with atoms having a momentum
$p=0$, the spontaneous emission produces ground state atoms with
an atomic momentum $p$ different from zero. These atoms in the $p\ne0$ ground state
constitute a thermal component separate from the condensate, to be
described through a sink state $|s\rangle$
whose atomic properties are equivalent to those of the $|g\rangle$ state. The
$|s\rangle$ state is filled by the spontaneous emission with
rate $\Gamma$ from the $|e\rangle$ state. The interaction of the
$|s\rangle$ state with the photoionization radiation is the same one of
the $|g\rangle$ state. However the photoionization from the
$|s\rangle$ state produces ions having an initial momentum $p\ne0$
and weakly interacting with the condensate. Therefore that photoionization
process is not relevant for our target of electrons/ions with a
small kinetic energy\cite{anderlini03} and it will not be considered into
our analysis.\\
\indent A laser with angular
frequency $\omega_{1}$, Rabi frequency $\Omega_{1}$ and
detuning $\delta=\omega_{1}-\omega_{\rm eg}$ acts on the $|g\rangle \to
|e\rangle$  transition, while a laser with
angular frequency $\omega_{2}$ and intensity $I_{2}$ acts on the $|e\rangle \to  |i\rangle$ transition.
The intensity of the first laser is
$I_{1}=\frac{\hbar \omega_{\rm eg}}{24 \pi \Gamma c^{3}}\Omega_{1}^{2}$. The ionization from the excited state
$|e\rangle$ is described by the $\sigma_{i}$ ionization cross
section for the laser with intensity $I_{i}$.  For rubidium the atomic parameters are
$\Gamma= 8.9\times 10^{6}$ s$^{-1}$,
$\sigma_{1}=4.7\times 10^{-24}$ m$^{-2}$
 for the 421 nm radiation and $\sigma_{\rm 2}=16.4\times 10^{-24}$
 m$^{-2}$ for the 1002 nm radiation\cite{courtade04}.\\
\indent The evolution of the atomic system interacting with the
laser radiation is given by the Optical Bloch Equations for the
atomic density matrix\cite{lambropoulos,shore,adler} on the
Hilbert space spanned by the four atomic states introduced above
\begin{eqnarray}
\dot{ \rho_{gg}}&=&\Omega_{1}{\rm Im}\tilde{\rho}_{eg} ,\label{RogEq}\\
\dot{\rho_{ee}}&=&-\Gamma \rho_{ee}-\Omega_{1}{\rm
Im}\tilde{\rho}_{eg}-\left(\sigma_{\rm 1}I_{1}+\sigma_{\rm
2}I_{2}\right)\rho_{ee},\\
\dot{\rho_{ss}}&=&\Gamma \rho_{ee},\\
\dot {\tilde{\rho}_{eg}}&=&-\frac{\Gamma}{2}
\tilde{\rho_{eg}}-i\delta\tilde{\rho}_{eg}-
\frac{\Omega_{1}}{2}\left(\rho_{ee}
-\rho_{gg}\right),\\
\dot {\rho_{ii}}&=&\left( \sigma_{1}I_{1}+ \sigma_{2}I_{2} \right)\rho_{ee}.
\label{density}
\end{eqnarray}
with the additional condition $\rho_{\rm gg}+\rho_{\rm
ee}+\rho_{\rm ii}+\rho_{\rm ss}=1$, where $\tilde{\rho}_{\rm eg}$
denotes the interaction representation of ${\rho}_{\rm eg}$. We
have not included into the density matrix equations the excitation
and ionization of the sink state $|s\rangle$ by the two lasers,
because those processes do not modify the number of atoms lost or
ionized from the condensate phase. The numerical results reported
in the following are valid not only for the simple three-level
scheme of Fig. 1, but also for the real rubidium level scheme,
with the excited state occupation spontaneously decaying to other
intermediate states before reaching the final sink
state\cite{courtade04}. Both the ions and the losses into the sink
states deriving from the condensate component are equivalent in
the two cases.

 The efficiency of the ionization will be measured
through the ratio $r$ between the atomic population in the sink
state and the atomic population  in the ionic state, after the end
of the laser interaction
\begin{equation}
r=\frac{\int_{0}^{\widetilde{\tau}}\rho_{\rm
ss}(t)dt}{\int_{0}^{\widetilde{\tau}}\rho_{\rm ii}(t)dt},
\label{efficiency-eq}
\end{equation}
where the time $\widetilde{\tau}$ is longer than the laser pulse
length $\tau$ in order to include also the spontaneous emission
losses towards the sink state at the end of the laser-atom
interaction. Notice that in this analysis a large (small) value of
$r$ corresponds to a low (high) ionization efficiency.\\
\indent For any temporal shape of the applied laser pulses, a key
parameter for the photoionization efficiency is the laser detuning
$\delta$ from the $|g\rangle \to |e\rangle$ transition. For
decreasing detunings the two-photoionization process becomes more
efficient owing to the large excitation of the intermediate
$|e\rangle$ state; at the same time, however, also the spontaneous
emission becomes more efficient. As a consequence, for decreasing
detunings the loss rate towards the $|s\rangle$ state usually
increases faster than the ionization rate, resulting in the
decrease of efficiency for the condensate atoms photoionization.\\
\indent

\begin{figure}
 \includegraphics[scale=0.7]{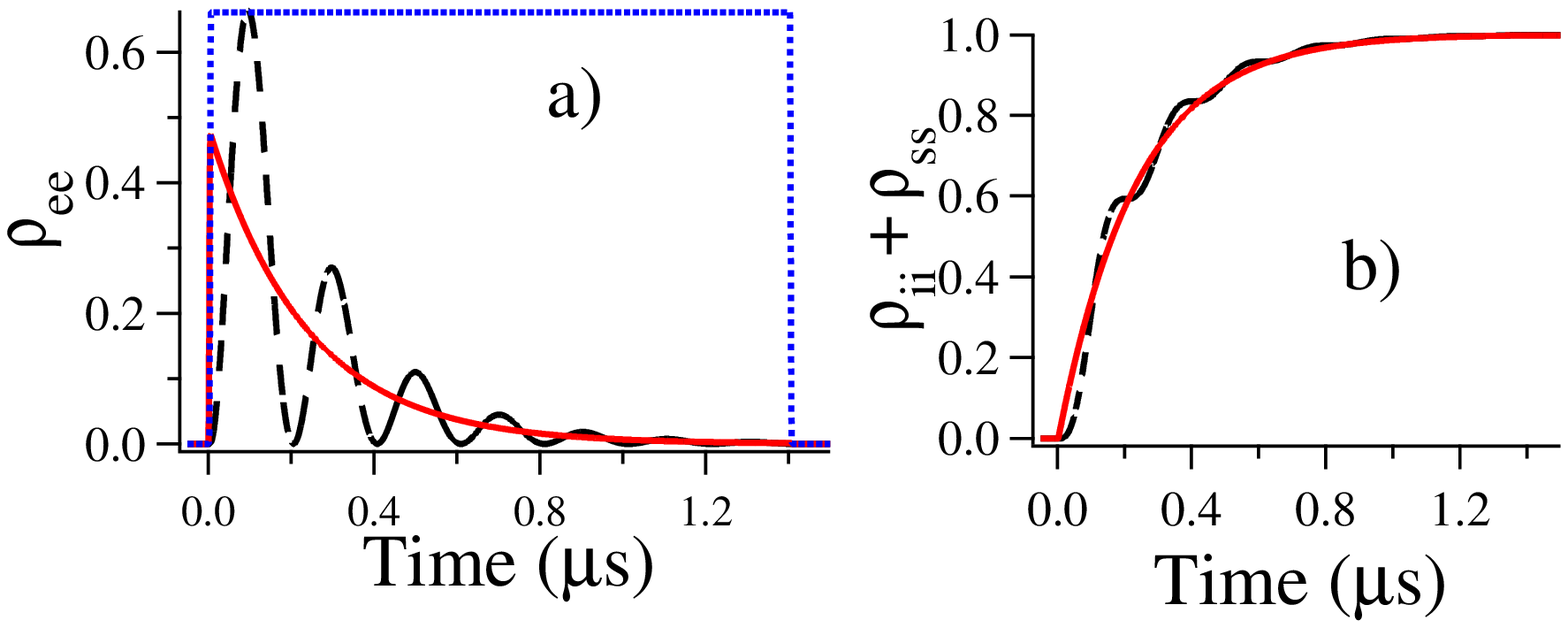}\\
 \caption{Comparison between the pertubative solution of
 Eqs.(\ref{longpulse-eq}) (solid lines) and the numerical solution of
Eqs. (\ref{RogEq}-\ref{density}) (dashed lines)  for the laser pulse
shape shown as dotted line in (a). In (a) fraction
$\rho_{\rm ee}$, in (b) sum of ionized
fraction $\rho_{\rm ii}$ and of sink-state fraction $\rho_{\rm
ss}$.  Laser
parameters $I_{1}=0.66$ Wcm$^{-2}$, corresponding to $\Omega_{1}/2
\pi=5.0 $
 MHz, $I_{2}=2.0\times 10^{3}$ Wcm$^{-2}$.}
 \label{longpulse-fig}
\end{figure}

\section{Long laser pulses}
During the simultaneous application of the two lasers, the atomic
population decays from the excited state $|e\rangle$ and is lost
from the system with total rate equal to $\Gamma_{\rm L}=\Gamma+
\sigma_{1}I_{1}+ \sigma_{2}I_{2}$. Therefore, at every time $t$
the total population in the $|g\rangle$ and in the $|e\rangle$
states decays with a time dependent loss rate equal to
$\Gamma_{\rm L} \rho_{ee}(t)$. In the case in which $\Gamma,
\sigma_{\rm i} I_{\rm i} \ll \Omega_{1}$ the ionized fraction of
population and the fraction lost by spontaneous emission can be
estimated by treating the losses as a perturbation of the steady
state solution for the Eqs.(\ref{RogEq}-\ref{density}) restricted
to the $[|g\rangle,|e\rangle]$ two-level system. This perturbative
solution is
\begin{equation}
\rho_{\rm ee}^{\rm
pert}(t)=\frac{\Omega_{1}^{2}}{2\Omega_{1}^{2}+\Gamma_{\rm
L}^2+4\delta^{2}}\,
\exp\left[-\frac{\Omega_{1}^{2}}{2\Omega_{1}^{2}+\Gamma_{\rm
L}^2+4\delta^{2}}\left(\Gamma
+\sigma_{1}I_{1}+\sigma_{2}I_{2}\right)t\right]
\end{equation}
Supposing the two lasers applied for a time $\tau$ the ionic and sink state
occupations may be written as
\begin{mathletters}
\begin{eqnarray}
 \rho_{\rm ii}(\tau)&=&\left(\sigma_{\rm 1} I_{\rm 1}+\sigma_{\rm 2} I_{\rm
 2}\right)\int_0^{\tau}\rho_{\rm
ee}^{\rm pert}(t)dt=\frac{\sigma_{\rm 1} I_{\rm
1}+\sigma_{2}I_{2}}{\Gamma_{\rm L}}
\left(1-e^{-\frac{\tau}{\tau_{\rm
L}}}\right),\nonumber \\
\rho_{\rm ss}(\tau) &=&\Gamma\int_0^{\tau}\rho_{\rm ee}^{\rm
pert}(t) dt= \frac{\Gamma}{\Gamma_{\rm L}}
\left(1-e^{-\frac{\tau}{\tau_{\rm L}}}\right).
\label{longpulse-eq}
\end{eqnarray}
\end{mathletters}
where we have introduced the timescale
\begin{equation}
\tau_{\rm L}=\frac{2\Omega_{1}^{2}+\Gamma^{2}_{\rm
L}+4\delta^{2}}{\Omega_{1}^{2}}\frac{1}{\Gamma+\sigma_{1}I_{1}+\sigma_{2}I_{2}}
\end{equation}
Therefore all the population is lost from the three-level system
if the duration $\tau$ of the pulses is much longer than
$\tau_{\rm L}$. In this limiting case, with no atom remaining in
the $|g\rangle$ state, the efficiency of the process is given by
 \begin{equation}
r_{\infty}=\lim_{\tau \to \infty}\frac{\rho_{\rm ss}(\tau)}{\rho_{\rm
ii}(\tau)}= f.
\end{equation}
with the $f$ parameter defined by
\begin{equation}
f=\frac{\Gamma}{\sigma_{\rm 1}I_{\rm 1}+ \sigma_{\rm 2}I_{\rm 2}}.
\label{fraction}
\end{equation}
For instance, at laser intensity $I_{1}= 0.66$W cm$^{-2}$
corresponding to a Rabi frequency $\Omega_{1}/(2 \pi)= 5$ MHz and
laser intensity $I_{2}=2000$ W cm$^{-2}$, the $f$ parameter is
equal to 53.8.\\

\section{pulsed excitation}
As demonstrated in refs \cite{vitanov02,halfmann}, under the
interaction with coherent pulsed radiation the atomic excitation
presents different properties than in measurements performed under
the cw excitation regime. In the regime of pulsed excitation, the
time evolution of populations and coherences becomes very
important. For calculating the ionization efficiency it is
necessary to perform the numerical solution of the density matrix
equations (\ref{RogEq}-\ref{density}). In fact, it should be taken
into account that the actual occupation $\rho_{\rm ee}$ of the
excited state is characterized by Rabi oscillations with effective
Rabi frequency $\Omega^{\rm eff}_{1}
=\sqrt{\Omega^{2}_{1}+\delta^{2}}$. Because the ionization depends
on the occupation of that state, the atomic ionized fraction
increases in time with oscillations at the same frequency.\\
\indent Fig. \ref{longpulse-fig} compares the exact solution of
the density matrix equations and the perturbation solution  given
by Eqs.(\ref{longpulse-eq}). The average ionized and the lost
fraction are well described by the perturbative solution. However,
due to the oscillatory evolution of the excited state population,
the numerical solution of Eqs. (\ref{RogEq}-\ref{density}) is
required in order to obtain the correct value of the ionization
efficiency parameter
$r$.\\
\indent Whenever $\tau\leq\tau_{\rm L}$, while no more ionization
occurs after the duration $\tau$ of the pulse and the total
ionized fraction remains equal to $\rho_{\rm ii}(\tau)$, a finite
fraction of population $\rho_{\rm ee}(\tau)$ remains in the
excited state at the end of the atom-laser interaction and decays
by spontaneous emission into the sink state. Therefore for a pulse
of duration $\tau$ the ionization efficiency ratio of Eq.
\ref{efficiency-eq} is given by
\begin{equation}
r(\tau)=\frac{\int_0^{\tau}\Gamma\rho_{\rm ee}(t)dt+\rho_{\rm
ee}(\tau)}{\int_0^{\tau}(\sigma_{\rm 1}I_{\rm 1}+ \sigma_{\rm
2}I_{\rm 2})\rho_{\rm ee}(t)dt}= {f+\frac{\rho_{\rm
ee}(\tau)}{\rho_{\rm ii}(\tau)}}. \label{rateefficiency}
\end{equation}
In general $r>f$ and the maximum efficiency of the ionization
process corresponds to the minimum value $r_{\infty}=f$.
\begin{figure}
 \includegraphics[scale=0.6]{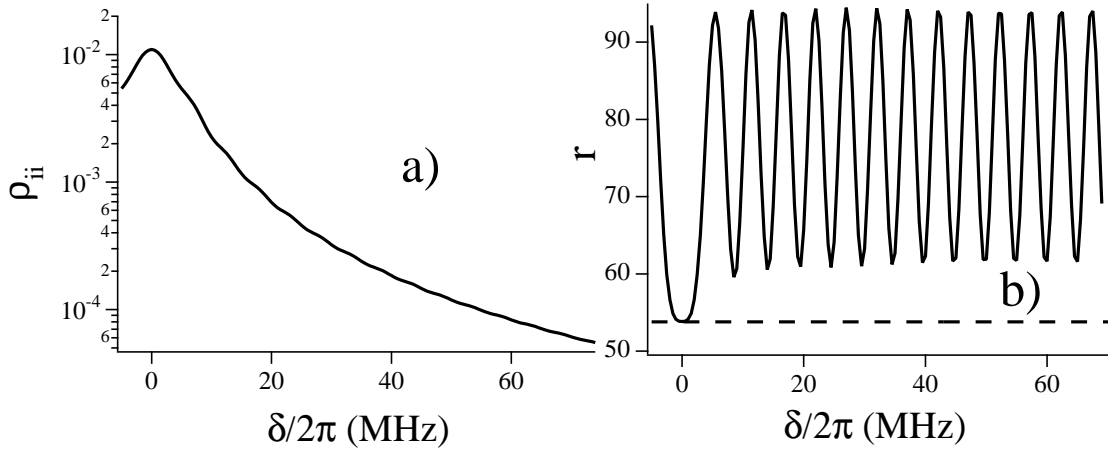}\\
 \caption{In (a) ionized fraction $\rho_{\rm ii}$ and in (b),
 continuous line, parameter $r$ of the efficiency ratio versus the blue laser detuning
 $\delta$ calculated through the numerical solution of the
 density matrix equations. The dashed line reports
 the limiting value given by the $f$ parameter of Eq. \ref{fraction}. Laser parameters
 $I_{1}=0.66$ Wcm$^{-2}$, corresponding to $\Omega_{1}/2 \pi=5.0 $
 MHz, $I_{2}=2.0\times 10^{3}$ Wcm$^{-2}$, interaction time $\tau =0.2/\mu$s.}
 \label{efficiency}
 \end{figure}

The oscillating behavior of the ionization process appears also
measuring the ionization as function of the detuning $\delta$.
Such a behavior is shown by the results of Fig. \ref{efficiency}
where the ionized fraction $\rho_{\rm ii}$ and the efficiency
ratio $r$ are plotted as a function of the laser detuning $\delta$
from the $|e\rangle$ excited state, at fixed interaction time
$\tau$. Owing to the choice $\tau=2 \pi/\Omega_{1}$, $\rho_{\rm
ee}(\tau)=0$, so that $r$ has a minimum (and the efficiency has a
maximum) at zero detuning. At larger values of the detunings
$\rho_{\rm ii}$ decreases with oscillations, and $\rho_{\rm
ee}(\tau)$ assumes values dependent on the effective frequency of
the Rabi oscillations. The role of the Rabi oscillations is more
clear on the plot of Fig. \ref{efficiency}(b) for the efficiency
ratio $r$. \\
\indent For fixed pulse duration the oscillating time dependence
of the excited state population with detuning-dependent effective
Rabi frequency $\Omega^{\rm eff}_{1}$ produces an oscillating
dependence of the ionized and of the lost fractions, and thus also
of the ionization efficiency, as a function of the detuning
$\delta/2 \pi$. In particular, the large oscillation of the
ionization efficiency is determined by the oscillating amount of
population remaining in the excited state $|e\rangle$ at the end
of the laser pulse and decaying totally by spontaneous emission
into the sink state $|s\rangle$. As a consequece, $r$ assumes
oscillating values which are always larger than the $f$ value of
Eq. (\ref{fraction}), and can be equal to that value only for
isolated detunings where the effective Rabi frequency $\Omega^{\rm
eff}_{1}$ produces a 2$\pi$ pulse for the
applied duration of the atom-laser interaction.\\
\begin{figure}
\includegraphics[scale=0.6]{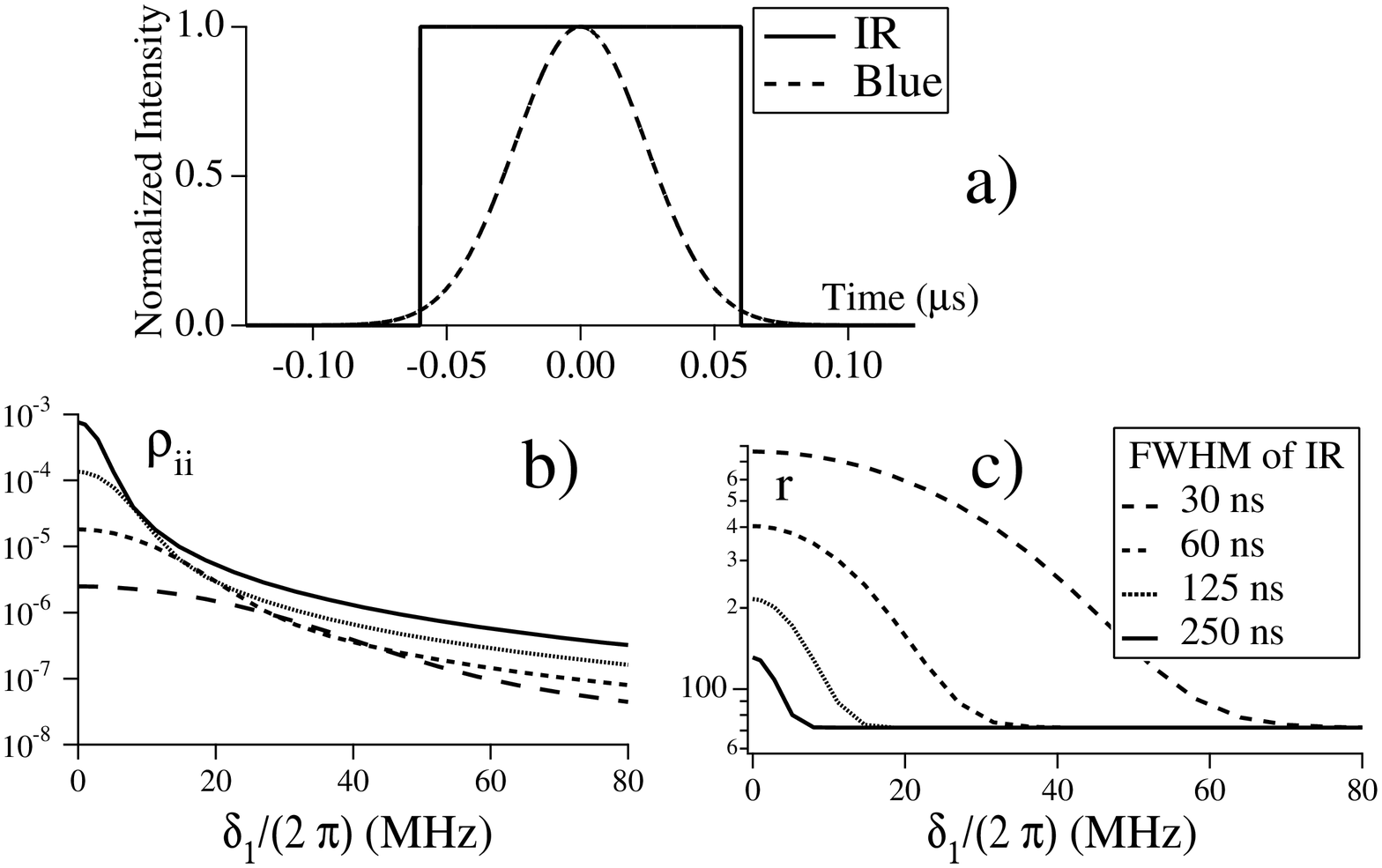}\\
\caption{Ionization with coherent population return. In a)
temporal shapes of the blue/IR laser for a 125 ns time length of
the IR pulse.  In b) ionic occupation $\rho_{ii}$ and in c)
parameter $r$ describing the efficiency ratio versus laser
detuning $\delta/2 \pi$, in MHz. $\Omega_{1}/2 \pi =5.$ MHz,
$I_{2}= 1500$ W/cm$^{2}$, $f=71.7$. Results for different time
lengths of the IR pulse, 30, 60, 125 and 250 ns respectively, with
the blue pulse length scaled as shown in a) and discussed in the
text.} \label{pulsed}
\end{figure}\\
\indent The occupation of the excited state at the end of the
laser pulse can be annulled in the regime defined as coherent
population return\cite{vitanov02,halfmann}. For the coherent
excitation of two-level atomic systems subjected to light pulses
of duration $\tau$ and detuning $\delta$ from resonance, the
dynamics of the atomic population is adiabatic if the detuning is
significantly larger than the Fourier width of the pulse, {\it
i.e.}, if $\delta \gg 1/\tau$. In this case the dressed state
occupied at the end of the excitation is the same one as before
the excitation. For a smooth pulse at constant detuning, that
state coincides with the atomic ground state, so that even if the
excited state is populated during the application of the pulse,
all the population is adiabatically brought back to the ground
state at the end of the process. Therefore no population fraction
remains in the excited state, $\rho_{\rm ee}(\tau)=0$. This regime
of coherent population return does not produces any loss into the
sink state at the laser pulse end. Results showing the advantages
of this regime in the two-photon ionization are reported in Fig.
\ref{pulsed}.\\
\indent We considered several durations of pulses all of
them with the shape shown in Fig. \ref{pulsed}(a). For each
duration of the IR square pulse, the blue intensity had a Gaussian
profile such that the intensity decreased to $1/e^{3}$ of the peak
value in correspondence of the IR pulse start and end. In Fig.
\ref{pulsed} (b) and (c) the ionized fraction $\rho_{\rm ii}$ and
the parameter $r$ of the the ratio between the atomic fraction in
the sink state and the ionized fraction are shown, respectively,
as a function of the laser detuning $\delta$, for several
durations of the laser pulse and at a given blue Rabi frequency.
In all cases, the effects of adiabatic excitation are visible, because for
each pulse duration there is a minimum detuning necessary to
obtain the optimal ratio $r$. The value of the minimum detuning is
only slightly sensitive to the Rabi frequency of the transition,
because, as pointed out in \cite{vitanov02}, no power broadening
occurs in a scheme of coherent population return. The limiting
value of $r$ reached at large detunings corresponds to the minimum
value $f$ calculated at the laser intensities of the numerical
analysis.

\section{Rapid Adiabatic Passage}
\begin{figure}
\includegraphics[scale=0.6]{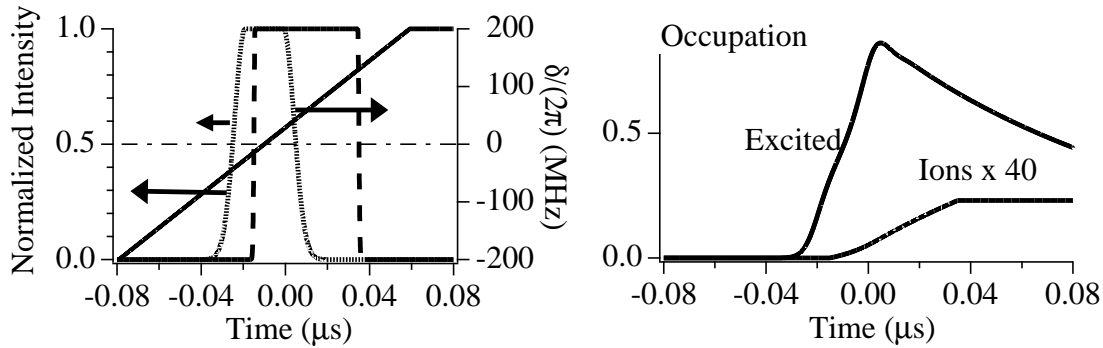}\\
\caption{Ionization using RAP from the ground state to the excited
state. In left plot the temporal shapes of the blue (dotted line)
and IR (dashed line) laser pulses and of the blue laser detuning
(continuous line) $\delta/2 \pi$, in MHz, at $\Omega_{1}/2
\pi=30.7$ MHz and $I_{2}= 2000$ W/cm$^{-2}$. In right plot time
evolution of the excited state occupation (solid line) and  of
ionized fraction excited state (dashed line).} \label{rap}
\end{figure}

Another excitation scheme allowing an efficient and robust
transfer of the population from the ground to the excited state is
the rapid adiabatic passage (RAP) technique\cite{abragam,guerin}.
This technique takes advantage of the population adiabatic
evolution in order to transfer almost the whole population from
the ground state to the excited state, blocking completely the
population return. RAP is produced by a time dependence of the
laser detuning, and for an efficient transfer
 the laser detuning changes from a very large and negative value to a
very large and positive value, or vice versa, during the
application of the light. The laser frequency experiences a chirp,
and the chirp duration should be shorter than the excited state
lifetime. Under these circumstances, the lowest energy adiabatic
state of the system coincides with the ground state before the
application of the light and with the excited state at the end of
the process. When this scheme is applied to the lower atomic
transition of the rubidium two-photon ionization, it allows to
transfer the ground state population to the $|6P\rangle$ state,
from where a pulse of IR light, with the time  sequence shown in
Fig. \ref{rap}(left-hand side), produces the ionization shown in
Fig. \ref{rap}(right-hand side). Because the IR pulse ionizes the
excited state atoms and also modifies the energy of the excited
state, for any temporal shape of the two laser pulses the largest
two-photon ionization is obtained for a particular delay of the IR
pulse with respect to the blue laser pulse. For the pulse shapes
of Fig. \ref{rap} the maximum ionization was obtained when the
center of the IR pulse was delayed by 0.02 $\mu$s with respect to
the center of the blue pulse. A linear chirp of the blue detuning
was examined, but the chirp exact shape had a minor influence on
the ionization efficiency, provided that the detuning difference
between initial and final value was larger than $\sim$100 MHz.
Suppressing the oscillatory behavior of the excited population,
the RAP led to an excited state occupation and an ionized fraction
larger than those obtained by pulses of the same shape but at
constant blue detuning. For the blue laser scanning of Fig.
\ref{rap} the ionization fraction was larger by a factor 2 than
produced at constant blue detuning (0.0057 against 0.0027). The
RAP technique is thus suitable for producing the maximum possible
fraction of ionized population within a laser pulse sequence of
short duration, during which the ratio between the atomic
population in the sink state and the atomic population in the
ionic state corresponds to the optimal value $f$. Due to the
suppression of the coherent population return, on the other hand,
a large fraction of population typically remains in the excited
state at the end of the laser pulse and decays into the sink state
thereafter. As a consequence, in RAP at long interaction times the
efficiency parameter $r$, calculated according to Eq.
\ref{efficiency-eq}, reaches values larger than those associated
to the adiabatic regime. For the parameters considered above the
efficiency ratio $r$ was 165 against the minimum value $f=53.8$.

\section{Conclusion}
We have examined a problem to be classified as a quantum
control one: reach a specific target, the ionization of the condensate,
minimizing the losses into a different output channel, the production of
thermal atoms. We have explored different two-photon ionization
schemes, producing different quantum controls for the atomic excited
state. However the quantum control reached for the two-photon ionization
is very poor. In effect the ionization process transferring atoms from
the excited state to the ionization state is governed by rate
equations, whence a classical response instead of a quantum one. Our
final result is that the maximum ionization efficiency to be reached
through the quantum control on the first excitation, depends on the
probabilities of two processes, the spontaneous decay rate and the
ionization rate. These rate are not modified within the quantum
control of our analysis. \\
\indent While the qualitative behavior of the ionized fraction
depends only on the properties of the laser pulse exciting the
first step of the ionization sequence, the absolute value of the
ionized fraction depends only on the intensity of the IR light
coupling the excited state to the continuum. An analogous
statement applies also to the efficiency ratio between the thermal
fraction and  the ionized fraction. The criteria we have derived
for the optimization of the two-photon ionization have a general
validity, whereas the absolute numbers we have discussed
correspond to the intensities available in the analyzed
experiment. The fraction of ionized atoms in these conditions is
much smaller than the number of photons scattered by each atom.
The application of the optimization criteria we have derived,
however, together with IR intensities larger than those we have
considered, will allow to reach experimental conditions leading to
an efficient large photoionization of Bose-Einstein condensates.

\section{Acknowledgements}
This research was supported
by the  INFM-Italy through the PhotonMatter Project, by the
MIUR-Italy through a PRIN Project, and by the European Commission
through the Cold Quantum-Gases Network, contract
HPRN-CT-2000-00125. E.A. thanks N. Vitanov for useful discussions.

\end{document}